\documentclass{aa}
\usepackage{graphicx,amsmath}
\usepackage{txfonts}

\usepackage{natbib} 
\bibpunct{(}{)}{;}{a}{}{,}
\begin{document}
\topmargin-3cm
 \title{The energy of waves in the photosphere and lower chromosphere: 1.~Velocity statistics}

   \author{C. Beck\inst{1} \and  E. Khomenko\inst{1} \and R. Rezaei\inst{2} \and M. Collados\inst{1}}
        
   \titlerunning{Energy of waves in the photosphere and lower chromosphere}
  \authorrunning{C. Beck et al.}  
\offprints{C. Beck}

   \institute{Instituto de Astrof\'{\i}sica de Canarias
     \and Kiepenheuer-Institut f\"ur Sonnenphysik
   }
 
\date{Received xxx; accepted xxx}

\abstract{Acoustic waves are one of the primary suspects besides magnetic fields for the chromospheric heating process to temperatures above radiative equilibrium (RE).}{We derived the mechanical wave energy as seen in line-core velocities  to obtain a measure of mechanical energy flux with height for a comparison with the energy requirements in a semi-empirical atmosphere model, the Harvard-Smithsonian reference atmosphere (HSRA).}{We analyzed a 1-hour time series and a large-area map of \ion{Ca}{II} H spectra on the traces of propagating waves. We analyzed the velocity statistics of several spectral lines in the wing of \ion{Ca}{II} H, and the line-core velocity of \ion{Ca}{II} H. We converted the velocity amplitudes into volume ($\propto \rho v^2$) and mass energy densities ($\propto v^2$). For comparison, we used the increase of internal energy ($\propto R \rho \Delta T$) necessary to lift a RE atmosphere to the HSRA temperature stratification.}{We find that the velocity amplitude grows in agreement with linear wave theory and thus slower with height than predicted from energy conservation. The mechanical energy of the waves above around $z\sim 500$ km is insufficient to maintain on a long-term average the chromospheric temperature rise in the semi-empirical HSRA model. The intensity variations of the Ca line core ($z\sim 1000$ km) can, however, be traced back to the velocity variations of the lowermost forming spectral line considered ($z\sim 250$ km).}{The chromospheric intensity, and hence, (radiation) temperature variations are seen to be induced by passing waves originating in the photosphere. The wave energy is found to be insufficient to maintain the temperature stratification of the semi-empirical HSRA model above 500 km. We will in a following paper of this series investigate the energy contained in the intensity variations to see if the semi-empirical model is appropriate for the spectra.}
\maketitle
\section{Introduction}
The chromospheric and coronal temperature rise above the photospheric
temperature requires a heating mechanism in addition to radiative energy
transport in the outer atmospheres of stars. The existence of chromospheres
and coronae on the Sun and other stars is a well established fact, but still
no generally accepted picture of the energy transport process responsible for
their formation could be derived, especially for the corona \citep[see e.g.][]{narain+ulmschneider1996}. 

The scenarios suggested for the chromospheric heating process can be divided in three main categories: 1. field-free heating by the propagation of (acoustic) waves in a gas plasma, 2. indirect magnetic heating in which magnetic field lines act as ``catalyst'' without being destroyed in the process, and finally, 3. direct magnetic heating in which the reconnection of magnetic field lines leads to a conversion of magnetic to thermal energy, thereby removing the field lines.

In the first case, acoustic waves are generated by material (convective)
motions as purely mechanical phenomenon \citep{musielak+etal1994}. The change of the atmospheric
properties in the vertical direction in a gravitationally stratified medium
leads to a ``steepening'' of vertically propagating waves producing a shock
front. The energy of the propagating wave is
dissipated in the ambient medium in and near the shock front, and hence,
mechanical energy can be transported over some distance before conversion to
thermal energy. Because of the stochastic nature of the excited acoustic
waves, the energy transport process would not be continuous \citep[e.g.][]{ulmschneider1971,ulmschneider+etal1978,carlsson+stein1997}.

For the second case of indirect magnetic heating requiring the presence of magnetic fields as a catalyst, two different scenarios are
possible in quiet Sun regions due to the specific topology of the magnetic fields: the
(more) vertical strong network fields \citep{solanki1993} and the (more)
inclined weaker internetwork (IN) fields
\citep{lites+etal1996,khomenko+etal2003,khomenko+etal2005,orozco+etal2007}. The network contains small-scale flux concentrations (some hundred kms) in the form of isolated flux tubes or flux sheets. These concentrated magnetic fields reside in the intergranular
lanes, because the interior of granules is swept free of magnetic fields by
the directed convective flows as long as the ratio of mechanical to magnetic
energy, $\beta \sim p_{gas}/B^2$, is above 1. The spatial extension of the flux
concentrations thus is strongly restricted in the photosphere by the
converging granular flows, but in the chromosphere the fields can expand
because of the reduced outside gas pressure and the lack of directed mass
flows \citep{hammer1987,keller+etal1990,solanki1993,steiner+etal1998}. This leads to a specific
topology with small-scale, mostly vertical fields in the photosphere that get more inclined
in the chromosphere and fill the whole atmospheric volume at some height. In
the layer of field expansion, a nearly horizontal canopy exists \citep{jones1985,schaffenberger+etal2006,lites+etal2008}. In this
topology, two possibilities exist to transfer energy from the photosphere
upwards in or near the network. The first option is that acoustic waves are generated {\em outside}
of the magnetic fields where some of them will propagate upwards as
before. The deposit of their mechanical energy is then not triggered by the
steepening in the gravitationally stratified medium, but the contact of the
propagating wave with (or transition into) the inclined magnetic field lines
\citep{chitre+davila1991,davila+chitre1991}. The second option is the
propagation of waves {\em inside} the magnetic fields themselves. These waves
can come in different modes (fast and slow
magneto-acoustic waves, Alf{\'v}en waves; see, e.g., Narain \& Ulmschneider
1996). They propagate along the field lines which with the topology of solar
quiet Sun fields would yield automatically a predominantly vertically
propagation. The deposit of energy then would be governed by the vertical
variation of atmospheric properties inside the field concentrations or mode
conversion processes. The source of the magnetic-field-related
waves would again be the convective motions of the surrounding granulation or p-mode oscillations that constantly buffet the magnetic field concentrations in the photosphere \citep{hindman+etal2008}. The
indirect magnetic heating by waves inside the fields would tend to affect a
larger volume in the chromosphere { than in the photosphere}, as the magnetic field lines directly connect a large chromospheric volume with its small photospheric counterpart. 
\begin{figure}
\centerline{\resizebox{6,5cm}{!}{\includegraphics{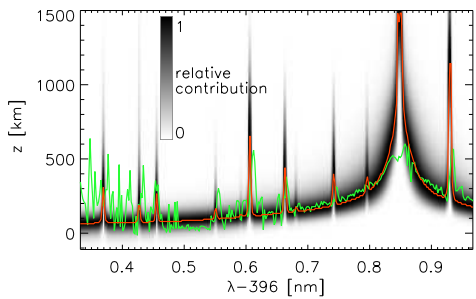}}}
\caption{Normalized intensity contribution function of \ion{Ca}{II} H in a LTE
  calculation. {\em Red}: center of contribution function. The {\em green} line was derived from phase differences \citep{beck+etal2008}.\label{response_geom}}
\end{figure}

For the third case of direct magnetic heating by reconnection, again two options are possible. Recent observations \citep{orozco+etal2007,lites+etal2008,steiner+etal2008} have shown that a
large amount of weak ($B<0.5$ kG) horizontal magnetic flux is present inside of granules
that seems to be swept upwards with the granular upflows. The existence of this
magnetic flux was already suggested to explain the large differences of
average magnetic flux between measurements using the Zeeman effect 
\citep[sensitive to cancellation of sub-resolution opposite-polarity fields, around 20 G;][]{khomenko+etal2005} and the Hanle effect \citep[60-100 G, insensitive to cancellation;][]{bianda+etal1999,trujillobueno+etal2004}. \citet{marian+etal2008}  invoked turbulent stochastically oriented fields to explain
the small amount of center-to-limb variation of linear and circular
polarization signal amplitudes in IN areas. \citet{lites+etal2008} recently increased the Zeeman-based estimate of the flux amount to around 60 G, using seeing-free data from the Hinode satellite. The interaction of the weak horizontal flux
with the magnetic canopy of the network or an increasing amount of reconnection between the 
horizontal fields with time could lead to an energy deposit in chromospheric
layers, where the magnetic field energy would have been produced previously in
the photosphere or below \citep[cf.][]{ishikawa+tsuneta2009}. The second option of magnetic heating would be the
reconnection of network fields triggered by the constant buffeting and
braiding of field lines due to the stochastic motion of the photospheric
foot points \citep{narain+ulmschneider1996,peter+etal2004}. 

Recently, \citet{fontenla+etal2008} have suggested the Farley-Buneman
instability as another possible source of chromospheric heating where
cross-field motions of ionized material lead to an energy deposit in a small
height layer. The horizontal magnetic fields again only act as a catalyst in
the process, but the motion of the ionized plasma is thought to come mainly
from convective motion rather than waves.

\begin{figure}
\centerline{\resizebox{6.5cm}{!}{\includegraphics{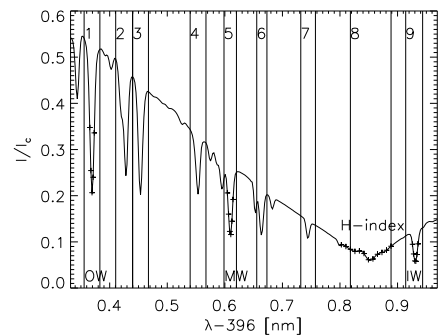}}}
\caption{Average \ion{Ca}{II} H spectrum with spectral lines 1 to 9 marked by {\em vertical lines} and intensity ranges OW, MW, H-index, and IW marked by pluses (+). \label{fig_lines}}
\end{figure}
\begin{table}
\caption{Line parameters.\label{tab_lines}}
\begin{tabular}{llllll}\cr\hline
Nr. & Ion  & $\lambda^1$& Exct. & log(gf) & Transition\cr
& & [nm] & pot. [eV] & &\cr\hline
1 & \ion{Cr}{I}  & 396.36900  &    2.544 &  0.66 & 5G 6.0- 5H 7.0  \cr 
2& \ion{Ti}{I}   & 396.42700  & 0.021  &-1.18   &3F 3.0- 3F 4.0  \cr
3& \ion{Fe}{I}   & 396.45530  &   2.839 & -1.55 &  3P 1.0- 5P 2.0 \cr
4& \ion{Fe}{I}   & 396.55088  &  3.241  &-1.778 & 5D 3.0- 5D 4.0 \cr
- & \ion{Fe}{I}   & 396.59152  &    2.425&  -3.550&  3P 1.0- 5D 1.0 \cr
5 & \ion{Fe}{I}   & 396.60617 &   1.605 & -1.64  & 3F 2.0- 3D 3.0 \cr
- & \ion{Fe}{II}  & 396.64260  &  1.724 & -5.525  &4P 0.5- 6D 0.5  \cr
6 & \ion{Fe}{I}   & 396.66300 &   3.212 & -0.360 & 5D 4.0- 5F 5.0  \cr
- & \ion{Fe}{I}   & 396.68100 &   3.301 & -1.949 & 3D 2.0- 3G 3.0  \cr
7 & \ion{Fe}{I}   & 396.74206  &   3.295 & -0.41  & 3H 4.0- 1G 3.0  \cr
- & \ion{Fe}{I}   & 396.79614  &  3.235  &-0.87  & 5D 3.0- 3G 4.0  \cr
8 & \ion{Ca}{II} & 396.84900 &   0.000 & -0.166&  2S 0.5- 2P 0.5 \cr 
9& \ion{Fe}{I} & 396.93000  &  1.485  & -0.42  &3F 4.0- 3F 3.0  \cr\hline
-&\ion{Fe}{I} & 630.15012   &  3.654 & -0.75 &  5P 2.0- 5D 2.0  \cr
-&\ion{Fe}{I} & 630.24936  &  3.686  &-1.236 & 5P 1.0- 5D 0.0  \cr
\end{tabular}\\$ $\\
1: wavelengths partly courtesy of J.Bruls, KIS (personal note), partly collected from the NIST Atomic Spectra Database \citep{martin+etal1998,reader+etal2002} and \citet{nave+etal1994}
\end{table}

From the observational point of view, any heating process related to
traveling waves is better accessible for detection than for example the direct
magnetic heating processes (granted that they exist). The best studied
chromospheric spectral lines, for which the importance of heating by waves, or
at least, the influence of waves has been definitely established, are \ion{Ca}{II} H and K at around 397 nm and 393 nm, respectively \citep{linsky+avrett1970}.
The \ion{Ca}{II} H line allows for example to trace the signature of waves at different atmospheric heights due to its broad line wings where only a region of around 100-200 km vertical extent contributes to the intensity at each wavelength \citep[cf.~ Fig.~\ref{response_geom} or Fig.~5 of ][]{rezaei+etal2008}. \citet{carlsson+stein1997} have demonstrated that the
(transient) bright grains seen in the core of H and K are due to acoustic
waves forming shocks. It is at present strongly debated if (high-frequency) acoustic
waves can supply enough energy to maintain a chromosphere
\citep[cf.][and the vivid discussion spawned by it]{fossum+carlsson2005}. The
energy of the acoustic waves is, however, usually derived indirectly from 
measurements of intensity variations observed in (broad-band) imaging. 

\citet{beck+etal2008} showed that the intensity of the emission peaks of \ion{Ca}{II} H has a clear phase relation to the intensity in the line wing for a large range of oscillation frequencies, i.e.~brightenings in the core are preceded by the same in the line wing \citep[cf.~also][]{liu1974}. They also showed that the Ca spectra of network locations show indications of transient events like the field-free\footnote{``Field-free'' is used here in the sense of ``no detectable circular polarization signal above noise level''.} IN regions, in addition to another contribution of steady emission. Thus, both magnetic network and field-free IN are significantly influenced by the presence of propagating waves.

In this contribution, we want to investigate the energy of these propagating waves by a detailed analysis of spectra of \ion{Ca}{II} H. The effect of the weak IN magnetic fields on the emission of \ion{Ca}{II} H was estimated to be negligible by \citet{rezaei+etal2007}. As the weak (or no) fields cover the largest part of the solar surface in the quiet Sun, a limitation to wave heating with neglect of direct magnetic heating processes seems reasonable. Note that the indirect magnetic heating processes are still included if they rely on propagating waves { that have a LOS component at disc centre}.
\begin{figure}
\centerline{\resizebox{7cm}{!}{\includegraphics{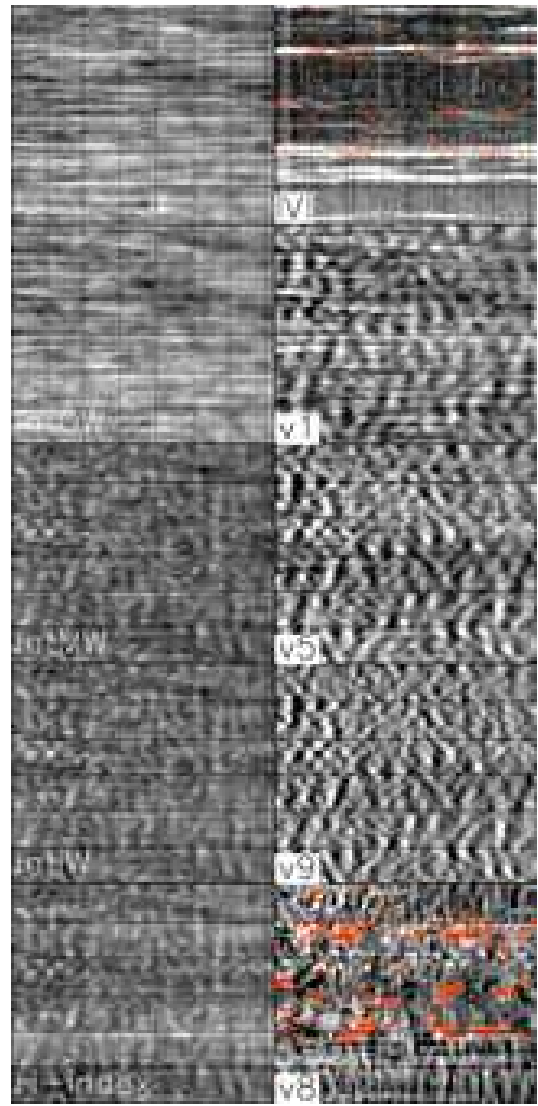}}}
\caption{Overview over the 1-hour time series. {\em Left column, top to
    bottom}: continuum intensity at 630 nm ($I_c$), intensity of outer line wing of \ion{Ca}{II} H (OW), middle wing (MW), inner wing (IW), H-index. {\em Right column, top to bottom}: integrated unsigned circular polarization, LOS velocity of line 1, 5, 9 (blends in line wing), 8 (\ion{Ca}{II} H line core). {\em Black/white} correspond to Doppler shifts of $\pm 1$ kms$^{-1}$ for line 1, to $\mp 1.5$ kms$^{-1}$ for lines 5 and 9, and to $\mp 5$ kms$^{-1}$ for line 8.  For the locations marked {\em red} no line-core velocity of \ion{Ca}{II} H could be defined; { the {\em red contour lines} mark these positions in the Stokes $V$ map. The {\em white contour lines} in v8 mark locations with strong $V$ signal.} The grid lines have a spacing of 10$^{\prime\prime}$ (y-axis) and 7 min (x-axis), respectively. \label{fig1}}
\end{figure}

After the description of the data sets used (Sect.~\ref{sec_obs}), we { determine the area fraction with signatures of magnetic fields in the chromosphere  in Sect.~\ref{mag_fields}, to check whether we can ignore field-related effects on the velocity statistics. We then investigate the relation between the line-core intensity and velocity of a set of spectral lines in Sect.~\ref{sec_intvelo}, to exclude a convective origin for the patterns. We study the statistical properties of the LOS velocities and the relations between the spectral lines of different formation height and compare the mechanical energy contained in the waves with requirements of a semi-empirical atmosphere model in Sect.~\ref{sec_velo}. The findings are summarized and discussed in Sect.~\ref{summ_disc}.} In a following paper of this series, we will investigate the energy in the intensity variations of the \ion{Ca}{II} H line for comparison with the (mechanical) energy contained in the LOS velocities.
\begin{figure}
\centerline{\resizebox{7.25cm}{!}{\includegraphics{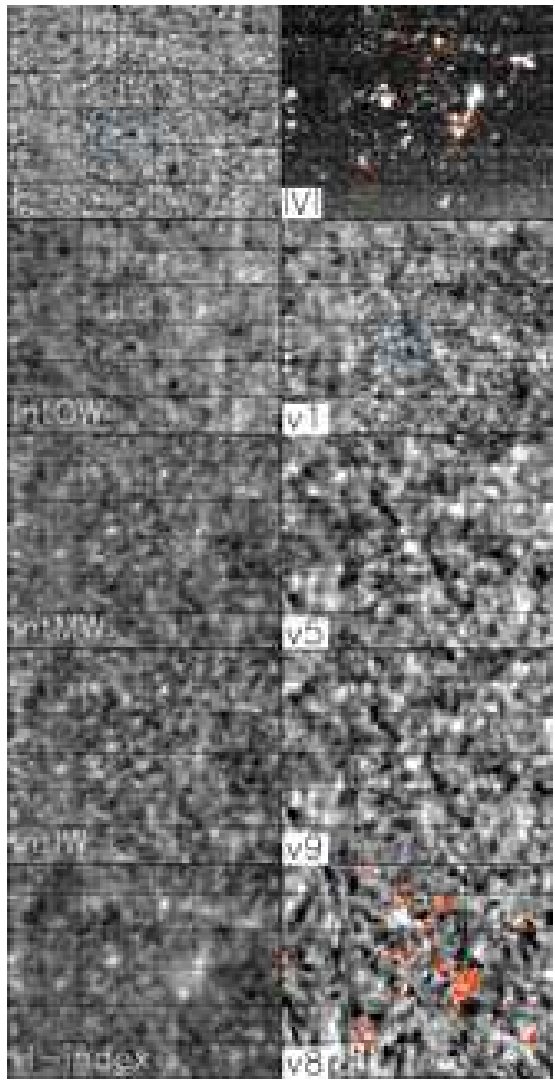}}}
\caption{Same as Fig.~\ref{fig1} for the large-area scan. The grid lines have a spacing of 10$^{\prime\prime}$ in both axes. { The {\em blue rectangle} marks an area of identical structure in $I_c$ and v1.}\label{fig2}}
\end{figure}
 \section{Observations, line selection and overview maps\label{sec_obs}}
We used two observations taken in the morning of 24/07/2006. The first is a
time series of around 1 hour duration (see Beck et al.~2008 for details, around 30000 spectra); the second is a large-area map of 75$^{\prime\prime} \times$ $\sim$ 70$^{\prime\prime}$ extent (also around 30000 spectra). The data sets were obtained with the POlarimetric LIttrow Spectrograph \citep[POLIS,][]{beck+etal2005b}. In both cases, the Stokes vector around 630 nm was obtained together with intensity profiles of \ion{Ca}{II} H. The slit width was 0\farcs5, as was the step width of scanning. Spatial sampling along the slit was 0\farcs29. The integration time was 3.3 sec per scan step for the time-series, giving a cadence of around 21 sec (repeated small maps of 4 steps). As in \citet{beck+etal2008}, we only used the last step with co-spatial spectra in 630 nm and \ion{Ca}{II} H. For the large-area map, the integration time per step was twice as large (6.6 sec). The data were reduced by the
respective set of routines for flat field and polarimetric correction
\citep{beck+etal2005b,beck+etal2005a}. The noise level in the polarization signal in continuum windows was around 1$\cdot 10^{-3}$ of the continuum intensity for the time-series and 8$\cdot 10^{-4}$ for the large-area map. Using (pseudo\footnote{For \ion{Ca}{II} H a spectral region in the line wing without a photospheric blend was used.}-)continuum intensity maps, the two wavelength ranges were aligned to be co-spatial. 

In the \ion{Ca}{II} H data, we identified several of the photospheric spectral line blends in the line wing. We used a subset of 9 spectral lines with a sufficient line depth in our study (cf.~Fig.~\ref{fig_lines}), together with the
two \ion{Fe}{I} lines at 630.15 nm and 630.25 nm in the second channel. Table \ref{tab_lines} lists all lines with line transition and adopted solar rest wavelengths. After deriving the  velocity statistics of all of them, it became clear, however, that the lines 1 to 4 (and 630.15 nm and 630.25 nm) had nearly identical properties, as had the lines 5 to 7. This can be related directly to their line depth, and hence, their similar formation height (cf.~Figs.~\ref{response_geom} and \ref{fig_lines} for lines 1 to 4 and 5 to 7). We then decided to used only lines 1, 5, 8, and 9 in the later evaluation. We attributed a ``line-core'' velocity to the displacement of the absorption core of \ion{Ca}{II} H, but caution that its interpretation as a velocity is less secure than for the other spectral lines. It also turned out that the statistics of both time-series and large-area map were as good as identical, so we usually used quantities averaged over both data sets if not noted otherwise.

Figures \ref{fig1} and \ref{fig2} show overview maps of the observations. The
{\em right column} shows the integrated unsigned Stokes V signal ({\em at top})  to mark locations with photospheric magnetic fields, and LOS velocities for the four chosen spectral lines. No reasonable velocities ($v\gg 10$ kms$^{-1}$) could be defined for the \ion{Ca}{II} H line core (line 8) on some locations because of the absence of a well-defined absorption core. These locations tend to be co-spatial to { or close-by to} strong photospheric magnetic fields and often show a very strong H$_{2v}$ emission peak with a plateau of constant intensity to the red of it, making a definition of velocity arbitrary. 
{ The maximum polarization degree
\begin{equation}
p= {\rm max} \sqrt{Q^2+U^2+V^2} \, ,
\end{equation}
can be used to estimate the presence and amount of photospheric magnetic flux. The fraction of pixels with a polarization degree $p$ above 1 \% of $I_c$ that indicate concentrated magnetic fields, and which correspond also to the bright patches in, e.g., the {\em top right panel} of  Fig.~\ref{fig2}, is around 10 \% (see \citet[][A\&A, in press]{beck+rezaei09} for a discussion of the dependence of $p$ on the spatial resolution)}. 
\section{Influence of chromospheric magnetic field topology on \ion{Ca}{II} H spectra \label{mag_fields}}
\subsection{{ Indirect magnetic field signature in intensity}}
{ The direct measurement of the chromospheric magnetic fields in the lower chromosphere has not been possible up to now for a number of reasons. The spectral lines like \ion{Ca}{II} H that form in a suited height range of the solar atmosphere are very deep and broad, leading to a low light level in observations of high spatial resolution. The magnetic fields are much weaker than in the photosphere; they induce only a weak polarization signal via the Zeeman effect. A detection of these weak polarization signals requires a high signal-to-noise ratio that is only possible with long integration times that, however, strongly reduce the spatial resolution again. Even if POLIS was designed for polarimetry in  630 nm and \ion{Ca}{II} H, the latter could only be done for solar features with strong  magnetic fields like pores or sunspots \citep[see][]{beck+etal2005b}. Lacking the direct detection of chromospheric magnetic fields, often the bright structures seen in imaging observations of chromospheric lines are taken as tracers of the magnetic field lines. 
\begin{figure*}
\sidecaption
\resizebox{12cm}{!}{\includegraphics{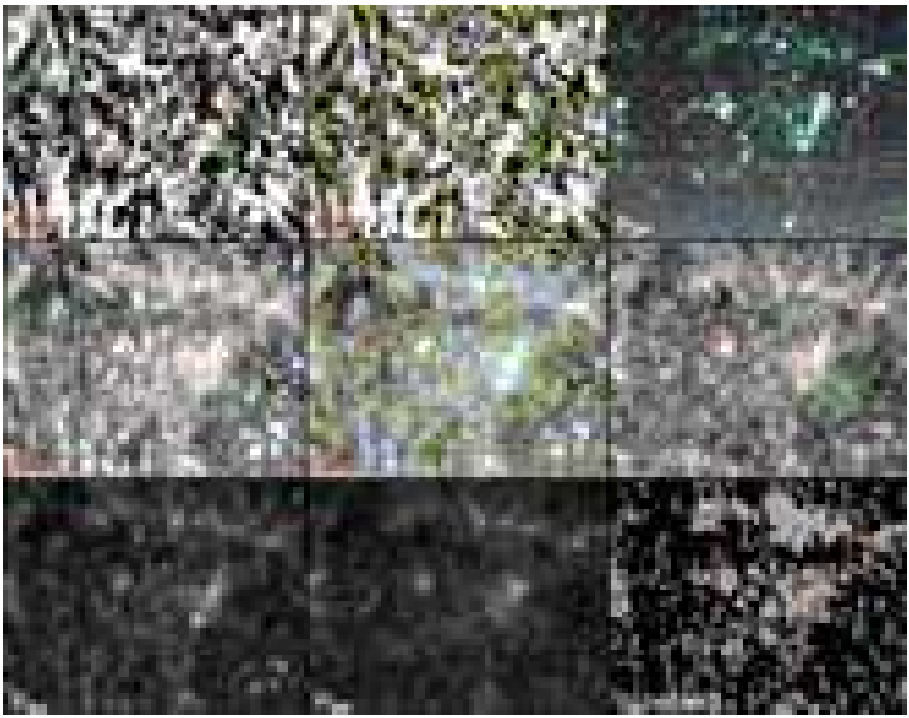}}
\caption{{ ({\em 1st column, top to bottom}): line-core velocity of \ion{Ca}{II} H, line-core intensity at 396.849 nm, intensity of H$_{\rm 2V}$ emission peak. ({\em 2nd column, bottom}): H$_{\rm 2R}$. The {\em upper two panels} are identical to 1st column, but with overlaid contours. ({\em 3rd column, top to bottom}): wavelength-integrated polarization degree $p_{tot}$, H-index, number of emission peaks in spectrum ({\em black/dark grey/light grey/white} correspond to 0/1/2/3 peaks). {\em Red contours} outline strong polarization signal, {\em blue contours} high line-core intensity, and {\em yellow contours} large blue shifts of the \ion{Ca}{II} H line core. The {\em green rectangle/parallelogram} denote the locations of fibril-like structures. {\em Green contours} in $p_{tot}$ mark profiles with two emission peaks.}\label{fig_mag1}}
\end{figure*}

The map of the wavelength-integrated H-index in Fig.~\ref{fig2} shows the strongest brightenings at locations with photospheric polarization signal. If one, however, reduces the averaged wavelength range significantly down to a few wavelength points, the intensity patterns change drastically. Figure \ref{fig_mag1} shows the FOV of the large-area map as seen in the very line core of \ion{Ca}{II} H ({\em middle left}). We averaged the spectra over a fixed wavelength range of $\pm 3$ pm around the line core at 396.849 nm. This intensity map shows several thin and elongated bright fibrils that originate from the patches with strong photospheric polarization signal. An identification of such fibrils with magnetic field lines is, however, not straightforward, if one compares with the line-core velocity map ({\em top left}). The intensity fibrils usually have a counterpart in the LOS velocity map, implying that the apparent increase of intensity in the map taken at a {\em fixed} wavelength is only due to a Doppler shift of the absorption core. We marked two prominent example of intensity fibrils with {\em green parallelograms}. The intensity fibrils have a counterpart in the velocity map with exactly the same morphology. This relation is not valid only for a few selected case, but throughout the whole FOV, as visualized by the {\em yellow contours} that mark strong blueshifts of the Ca line core. The shape of these velocity contours inside the {\em green rectangle} is complementary to the bright fibrils in the line-core intensity. To improve the visibility of the features, we displayed the same line-core velocity and line-core intensity maps twice in 1st and 2nd column, as the overlaid contour lines unavoidably tend to hide the fine-structure. \citet{leenaarts+etal2009} found a similar relation between velocity and intensity for the \ion{Ca}{II} IR line at 854 nm in observations and simulations.

The apparent intensity increase in the line-core intensity map does not extend far beyond the locations of large photospheric polarization signal ({\em blue} and {\em red contour lines}, respectively), whereas for the H-index the brightest parts are to first order co-spatial with the polarization signal ({\em middle right panel}). The H-index is not sensitive to the Doppler shifts producing the bright fibrils due to its extended wavelength range, contrary to the line-core intensity map. 
\subsection{Direct signature of magnetic fields in spectral shape}
Magnetic fields leave, however, also a more direct signature in the shape of the core of the \ion{Ca}{II} H line profiles. A spectral pattern with double-peaked emission on the locations of the H$_{\rm 2R}$ and H$_{\rm 2V}$ emission peaks is found almost exclusively together with photospheric magnetic fields \citep{liu+smith1972,rezaei+etal2007,beck+etal2008}. But even if both emission peaks are present at the same time in many cases, they usually show a pronounced asymmetry with higher intensities in H$_{\rm 2V}$. This asymmetry indicates the presence of mass flows \citep[e.g.,][]{heasley1975,rezaei+etal2007}. 

To determine the locations inside the FOV with a direct magnetic field signature, we have analyzed the shape of the \ion{Ca}{II} H profiles with a routine similar to the one used by \citet{rezaei+etal2008}; the routine searches for local intensity maxima in the spectrum. Appendix \ref{app_profiles} explains the procedure in more detail and shows several examples of its results on different profiles. With this routine the intensity of H$_{\rm 2V}$ or  H$_{\rm 2R}$ can only be determined when the peaks are present in the spectrum. Following \citet{rezaei+etal2007} we thus also determined the integrated intensity in two 27 pm wide wavelength bands located symmetrical around the line core that yield an estimate of H$_{\rm 2V}$ and H$_{\rm 2R}$ intensity for each profile. 

The {\em bottom row} of Fig.~\ref{fig_mag1} shows the results of both approaches that use the shape of the spectrum near the line core. The appearance of spectra with both H$_{\rm 2V}$ and H$_{\rm 2R}$ emission ({\em green contour lines} in the polarization degree map at {\em upper right}) is basically restricted to the locations with photospheric polarization signal, with an additional extent of about 5$^{\prime\prime}$ distance around the photospheric fields. In the rest of the FOV, either single peaked profiles or those without intensity reversal are prevailing. The area fraction of double-peaked profiles in the FOV is 13\%, slightly larger than the estimate of photospheric magnetic field coverage. The maps of H$_{\rm 2V}$ and H$_{\rm 2R}$ using the 27 pm wavelength bands are displayed in the {\em bottom row}, with an identical display range; they show that the H$_{\rm 2V}$ emission is usually stronger than H$_{\rm 2R}$, even on the locations with photospheric magnetic fields. For half of the profiles with two emission peaks the asymmetry between H$_{\rm 2V}$ and H$_{\rm 2R}$ is above 10 \%. As discussed by \citet{rezaei+etal2007} and \citet{beck+etal2008}, this is related to the fact that the Ca profiles even above photospheric magnetic fields contain a (dominant) contribution with the signature of acoustic shocks, indicated by a pronounced H$_{\rm 2V}$/H$_{\rm 2R}$ asymmetry \citep[][their Fig.~17]{beck+etal2008}.
\begin{figure*}
\centerline{\includegraphics{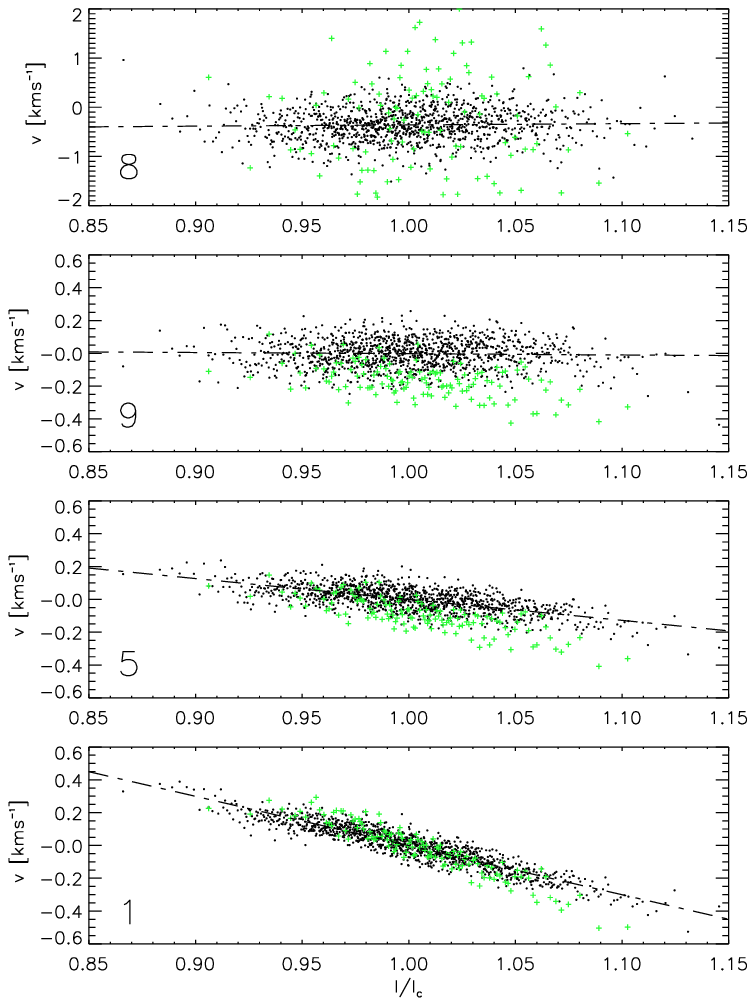}\includegraphics{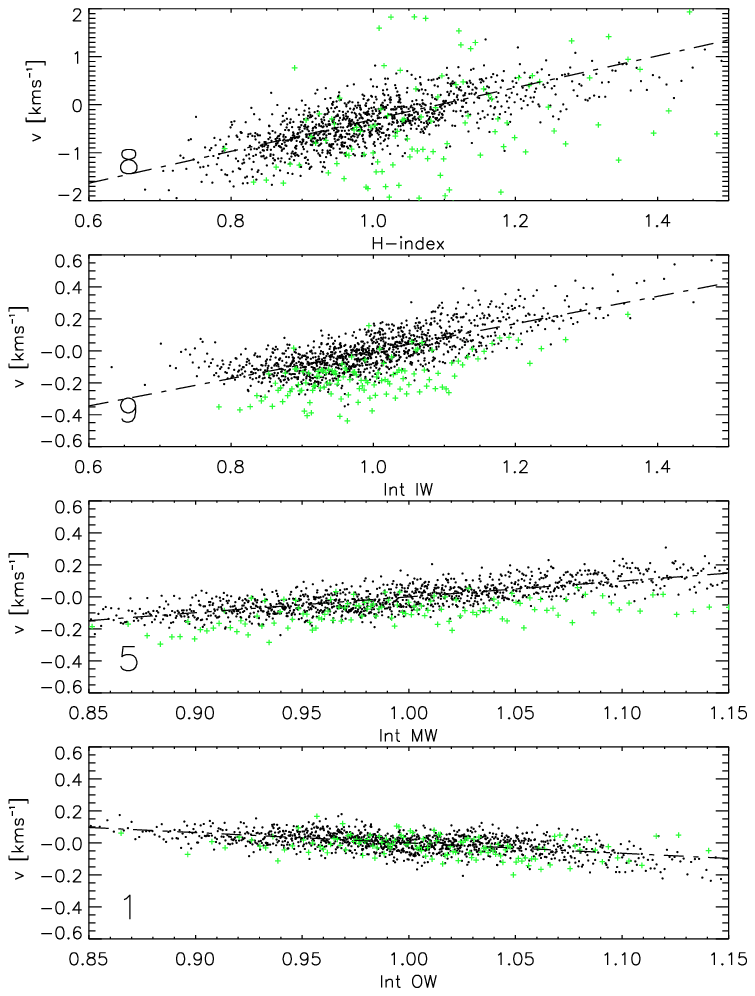}}
\caption{{\em Left column}: scatterplot of continuum intensity vs LOS
  velocities. {\em Dash-dotted lines} show a linear regression fit. The data
  has been binned. {\em Right column}: scatterplot of the LOS velocities vs
  their respective line-core intensities. { The {\em green pluses} show the same for locations with two emission peaks in \ion{Ca}{II} H.}\label{i_vs_v}}
\end{figure*}

In total, the area fraction with chromospheric magnetic field signature is small ($\sim$ 15 \%), half of the affected profiles show indications for the presence of flows as well, and we are interested in the velocity statistics of all type of waves regardless whether they are related to magnetic fields or not. We thus assume that the magnetic field topology can be neglected in the analysis of the velocity statistics of the full FOV. To investigate whether the chromospheric magnetic fields have any effect on the LOS velocities, we will use the statistics of all locations with a double-peaked emission in the large-area map ({\em light grey} area in {\em bottom right} of Fig.~\ref{fig_mag1}) later on for a comparison with the full FOV. These locations cover most of the photospheric polarization signal as well, and thus indicate magnetic fields at all heights. 

We would like to also point out that the bright fibrils may nonetheless actually still {\em outline} chromospheric magnetic field lines. The reason for their appearance may be, however, not the presence of the field lines alone, but rather the passage of a wave or mass flows along them. This corresponds to one of the indirect magnetic heating scenarios discussed in the introduction.}
\section{Intensity-velocity relation\label{sec_intvelo}}
To obtain the intensity maps (outer wing (OW), middle wing (MW) and inner wing (IW), H-index) in the {\em left columns} of Figs.~\ref{fig1} and
\ref{fig2} that should correspond to approximately the same formation height
as the LOS velocities, we averaged the intensity in the \ion{Ca}{II} H spectrum over the line cores of the respective lines (pluses in Fig.~\ref{fig_lines}). Even if the averaging extends over some spectral pixels, we will address the resulting intensity maps as the ``line-core intensity'' of the respective line in the following. The approach, however, only partly succeeded in the desired effect of producing intensity maps looking similar to the velocity maps. For the lowest forming spectral line (line 1, display range $\pm$ 1 kms$^{-1}$; Fig.~\ref{fig2}), the spatial structures in the velocity and the line-core or continuum intensity map do match: white in velocity (= blue-shifts) corresponds to an increased intensity, but for the other lines the relation between intensity and velocity is less clear. The display range was increased to $\pm 1.5$ kms$^{-1}$ for line 9 and $\pm 5$ kms$^{-1}$ for line 8. For all velocity maps besides v1 we reversed the velocity display range (white = red-shifts). 

In Fig.~\ref{i_vs_v} we tried to quantify the relation between intensity and velocities for two cases: velocity vs the continuum intensity $I_c$ ({\em left column}) and velocity vs the respective line-core intensity of the spectral line (pluses in Fig.~\ref{fig_lines}; {\em right column}). { The line-core intensities were normalized with their mean values. To improve the visibility of the dependence, the data were binned in the same way as in \citet{beck+etal2007}. We took the pairs of corresponding ($I, v$) points, sorted the intensity values into fifty equidistant bins $[I,I+\Delta I]$, and then used the average values $\overline{I} = \langle I \rangle _{[I,I+\Delta I]}$ and $\overline{v} = \langle v \rangle _{[I,I+\Delta I]}$ in each bin. }

For lines 8 and 9, no correlation with the continuum intensity remains, whereas line 5 and line 1 show a weak and strong anti-correlation, respectively, with the granulation pattern of $I_c$. Comparing the velocities to the respective line-core intensities, already line 5 shows a reversal of the relation, with red-shifts corresponding to increased intensities. { Restricting the FOV to only the locations with double-peaked emission gives essentially the same relationship between velocities and intensities ({\em green pluses}).} 

A second peculiarity of the velocity and intensity maps relates to the spatial scales. Whereas for the velocities the spatial scales of structures in the large-area map (Fig.~\ref{fig2}) increase to a more diffuse pattern, this trend is less prominent in the line-core intensity maps. Along the temporal axis of the time series (Fig.~\ref{fig1}), the velocity patterns do not turn diffuse with height. { This should be related to the frequency shift of the oscillations with height (cf.~Appendix \ref{appa}) that lead to a compression of the wave pattern in the temporal domain.}

In brief, both the intensity and the velocity maps other than for line 1 do not show any clear relation to the granulation pattern, implying that they are dominated by processes other than the convective motions. The spatial structures in the velocity maps show a typical spatial scale that is larger than individual granules and that could be part of the inverse granulation pattern \citep[e.g.][]{rutten+etal2004a}. These authors concluded that the inverse granulation pattern should be a mixture of convection reversal and a contribution from gravity waves whose relative importance increases with height. The little visual correspondence between velocity and intensity maps then again supports the claim that line 5 and those higher up are dominated processes unrelated to convection.
\begin{figure}
\centerline{\resizebox{6.5cm}{!}{\includegraphics{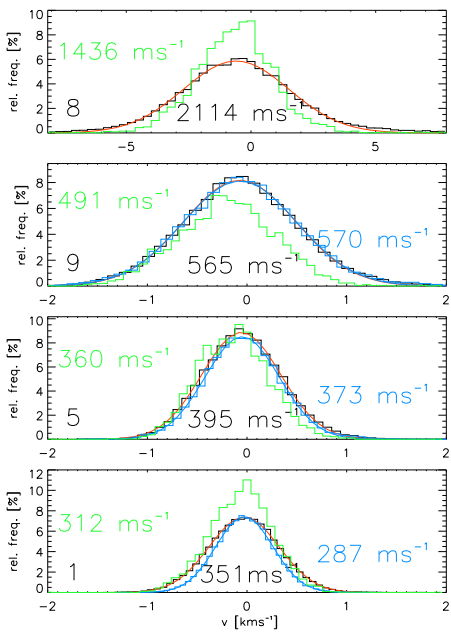}}}
\caption{Histograms of LOS velocities. Spectral line and width of the Gaussian fit ({\em red line}) are indicated in each plot. The {\em blue} curves and values {\em at right} refer to the Fourier-filtered velocity maps of the time-series; { the {\em green curves} and the values {\em at left} are for locations with two emission peaks in \ion{Ca}{II} H}.\label{fig_hist}}
\end{figure}
\begin{figure}
\centerline{\resizebox{6.5cm}{!}{\includegraphics{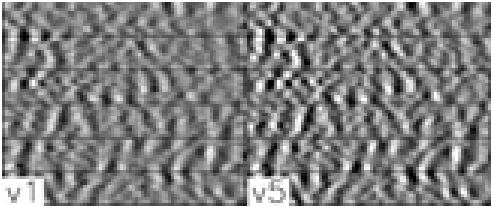}}}
\caption{Fourier-filtered velocities of line 1 and 5 of the time-series. The grid lines have a spacing of 10$^{\prime\prime}$ (y-axis) and 7 min (x-axis).\label{fourier_filt}}
\end{figure}
\begin{figure}
\centerline{\resizebox{8.cm}{!}{\includegraphics{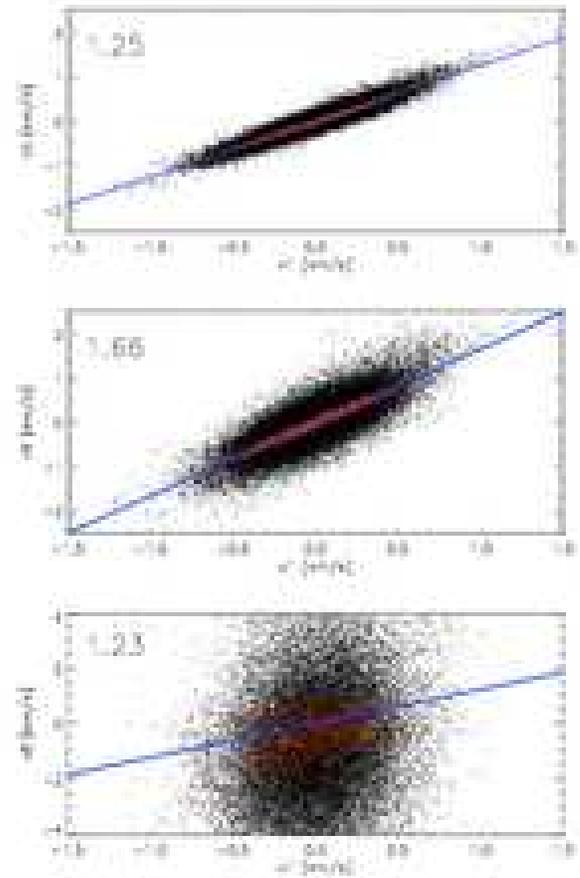}}}
\caption{Relation between co-spatial velocities of lines 1,5,9 and 8 in the time-series. {\em Top to bottom}: scatterplots of v1 vs v5, v1 vs v9, and v1 vs v8. {\em Black dots}: all data points; {\em red}: after binning. {\em Blue line}: linear regression to the binned data. The slope is given in the upper left corner in each scatterplot. \label{vel_exam}}
\end{figure}
\section{Statistics of LOS velocities with height\label{sec_velo}}
The histograms of the line-core velocities in Fig.~\ref{fig_hist} clearly show that the width of the velocity distribution increases with height, as the lines have been sorted according to their formation heights given in Table \ref{tab_ampl}. We made a Gaussian fit to the distributions to define the root-mean-square (rms) value  $\sigma_{vi}$ of velocity variations from the width of the distribution. For the data from the time-series, it is possible to remove the velocities induced by the convective motions by Fourier filtering. We took out all velocities corresponding to frequencies below 1.3 mHz ($\sim$760 sec period) from the velocity maps of line 1, 5, and 9. Figure \ref{fourier_filt} shows the resulting velocity maps for line 1 and 5. The removal of the low-frequency variations changes the distributions of velocities slightly by reducing their width by 60 (20) ms$^{-1}$ for line 1 (line 5). For line 9, which showed no correlation to the granulation pattern, the effect is negligible ({\em blue curves} in Fig.~\ref{fig_hist}). In the following, we only used the low-pass filtered velocity from the times-series for line 1 and 5, neglecting the large-area map for these lines, but used both data sets for lines 8 and 9 where the filtering is negligible. { Restricting the FOV again to only the locations with double-peaked emission ({\em green lines}) yields distributions with a slightly reduced width for all lines. We will not use the statistics for the magnetic locations further; the main difference to the full FOV is the reduced rms velocity value for \ion{Ca}{II} H, whereas for the other lines the reduction is less severe.}

{ Up to this point, three arguments show that the patterns in the velocity maps cannot be of convective origin:
\begin{itemize}
\item The low-forming lines have been Fourier-filtered for the low-frequency part which removes the granular contribution; only the filtered time-series is used for lines 1 and 5.
\item The velocity maps for lines 9 and 8 have no correlation to the granulation pattern.
\item All lines are dominated by oscillatory behavior with recurrent positive and negative velocities with periods shorter than granular time scales (cp.~Figs.~\ref{fig1} and \ref{fourier_filt}).
\end{itemize}
The patterns thus should be related to waves of some kind; using the velocity maps of the time-series shows that they are propagating in height as well.
} 

\begin{figure}
\resizebox{8.7cm}{!}{\includegraphics{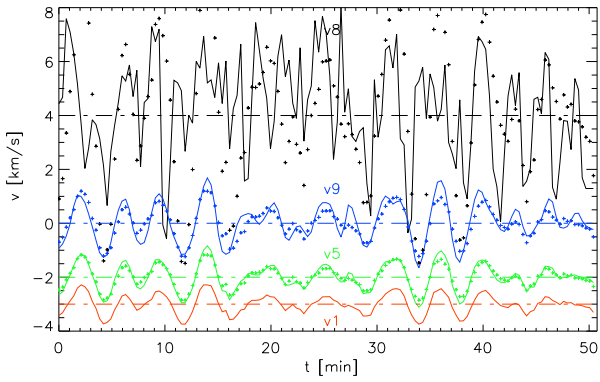}}
\caption{Example of LOS velocities at a fixed position along the slit in the time-series. {\em Black, blue, green} and {\em red} lines show the velocity of lines 8,9,5, and 1, displaced by +4,0,-2,-3 kms$^{-1}$. The {\em dashed} lines mark the respective zero velocity. The {\em colored crosses} result from multiplying v1 with the coefficients of Fig.~\ref{vel_exam} for v5 and v9, and the ratio of the rms velocities for v8, respectively.\label{vel_exam_a}}
\end{figure}

The question of the wave energy during the propagation through the atmosphere is related to the growth of the velocity amplitude of the vertical LOS velocities with height that has to compensate the reduction of gas density. Figure \ref{vel_exam} compares co-spatial and co-temporal velocities in the time-series. To obtain the velocities co-temporal to line 1, the velocity maps of line 5, 9, and 8 were shifted in time by 5 sec, 10 sec and 59 sec, respectively. Figure \ref{vel_exam_a} shows that with these temporal shifts, the velocities of the lines at one location along the slit are well in phase for lines 1, 5, and 9. For the Ca line core, the agreement of phase is worse, but for most of the large-amplitude velocity excursions the 1-min shift gives a good agreement (e.g., at t=5,32, or 40 min). For the co-spatial and co-temporal velocity maps, we did scatterplots vs the velocity of line 1 (Fig.~\ref{vel_exam}). The different lines show a strong correlation with the lowermost forming line 1 with an ever increasing velocity amplitude ratio. The ratio was determined by a linear regression fit to binned values of the scatterplot ({\em red dots}). The obtained ratio  was then used to scale up the velocity of line 1 to the velocities to be expected for the other lines ({\em colored crosses} in Fig.~\ref{vel_exam_a}). For line 5 and line 9, the usage of a constant scaling factor reproduces the actual observed velocity amplitudes well, regardless of the period of the velocity oscillations. For the Ca line core, the scaling coefficient from the scatterplot ($\sim$1) leads to much too small predicted velocities; we thus scaled line 1 up with the ratio of the rms velocities instead ($\sim$7). This reproduces only the range of observed Ca line core velocities; no close match in phase or amplitude at a given time is achieved. In brief, all velocity oscillations of line 1 appear in the other lines with both a time-lag and and an amplitude increase. 

\begin{figure}
\centerline{\resizebox{7.cm}{!}{\includegraphics{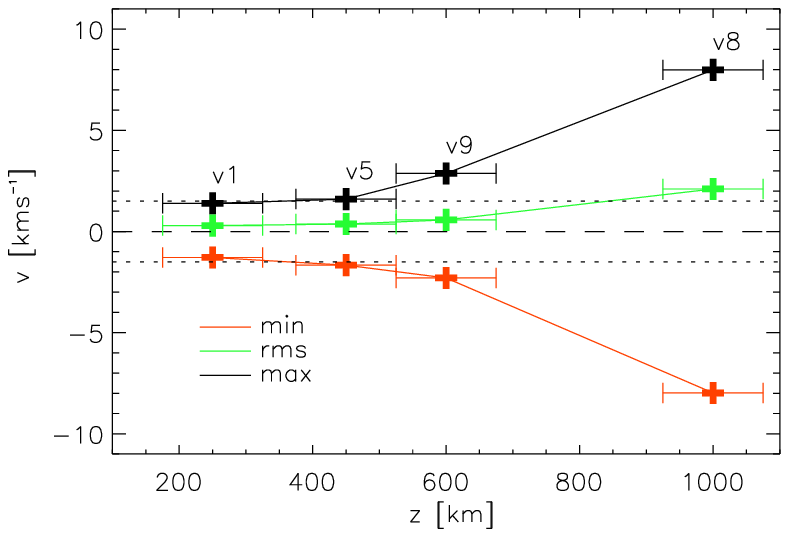}}}
\centerline{\resizebox{7.cm}{!}{\includegraphics{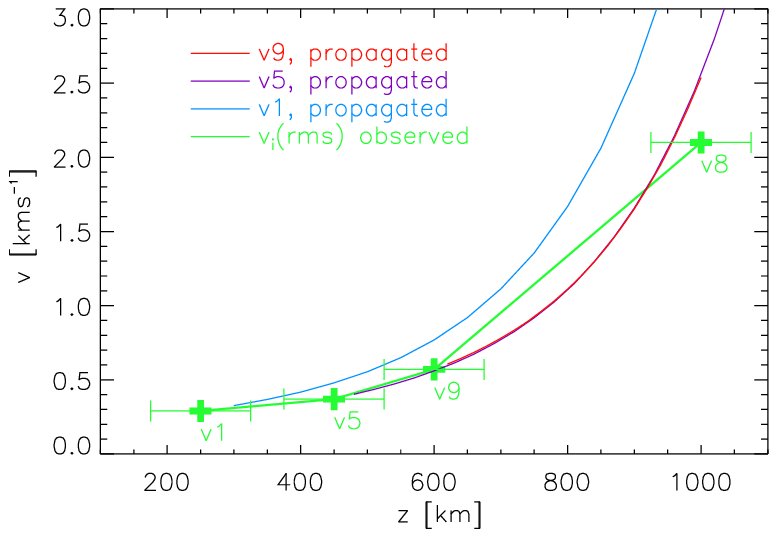}}}
\caption{{\em Top}: statistical properties of LOS velocities: minimum, maximum and rms value ({\em red, black, green}). The {\em horizontal dotted lines} are at $\pm$ 1.5 kms$^{-1}$, the {\em dashed} at zero. The {\em horizontal error bars} correspond to a $\pm 75$ km error in the assumed formation heights. {\em Bottom}: comparison between observed rms velocities ({\em green}) and those from a propagating of velocities in height using linear wave theory. {\em Blue line}: propagation of v1;  {\em purple line}: same for v5,  {\em red line}: same for v9.\label{velo_range}}
\end{figure}

The statistical properties of the lines are summarized in the {\em upper part} of Fig.~\ref{velo_range}: the extreme velocities increase from $\pm$ 1.5 kms$^{-1}$ in line 1 to above $\pm$ 8 kms$^{-1}$ for the line core of \ion{Ca}{II} H, the rms value increases from 0.3 kms$^{-1}$ to 2 kms$^{-1}$. 
\begin{table}
\caption{Line-core formation height ({\em 2nd column}) and velocity amplitude ratios of {\em left to right} observed rms velocities, velocity of v1 propagated in linear wave theory, same for v5, and from the fit to the scatterplots.\label{tab_ampl}}
\begin{tabular}{llllll}
line &z [km] & $\sigma_{vj}/\sigma_{v1}$ &$\sigma_{v1,prop.}/\sigma_{v1}$& $\sigma_{v5,prop.}/\sigma_{v1}$  & fit\cr \hline  %$\sigma_{v1,prop.}/\sigma_{v1}$
1 & 250 &1 & 1  & - & 1 \cr   %1
5 & 450 &1.28  &1.65& 1.28& 1.25 \cr  % 1.65& 
9 & 600 &1.99 & 2.65& 1.93& 1.66 \cr % 2.65&
8 &  1000 &7.37  & 8.75& 8.83& 1.23 \cr %8.75&
\end{tabular}
\end{table}

To investigate the consistency of the observed velocities, we propagated the velocities in height using the linear perturbation approximation for wave propagation. We used the Fourier transform of the velocity maps of lines 1, 5 and 9, and propagated the velocities upwards using \citep{mihalas+mihalas1984}
\begin{equation}
v(z+\Delta z,\omega) = v_i(z,\omega)\cdot \exp\left( \left(\frac{1}{2H} - \frac{\sqrt{\omega_c^2-\omega^2}}{c_s}\right)\cdot \Delta z \right) \;. \label{eq_prop}
\end{equation}
A constant scale height of $H$ = 100 km was used, the sound speed $c_s$ was set to 7 kms$^{-1}$ and an acoustic cutoff frequency of $\nu_c = 5$ mHz was used to calculate $\omega_c=2\pi \nu_c$. For each height $\Delta z$ we then calculated the velocity histograms and rms velocities like for the observations (see Appendix \ref{appa} for more details). The {\em lower part} of Fig.~\ref{velo_range} compares the observed and synthetic rms velocities from the propagation. The propagation of v1 leads to rms velocities larger than the observed ones for the other lines, but if v5 is propagated with Eq.~(\ref{eq_prop}) a good match to the observed values of line 9 and 8 is obtained. This indicates again that the velocities from line 5 and higher up are dominated by waves. Table \ref{tab_ampl} compares the observed velocity amplitude ratio between line 1 and the other lines with that of the propagated velocities and the linear fit to the scatterplots of Fig.~\ref{vel_exam}. Besides the values from the propagation of v1, the other values agree well that the velocity amplitude increases on average by (1.3, $\sim$2, $\sim$ 8) for $z$=(450, 600, 1000) km.  
\begin{figure}
\centerline{\resizebox{6.5cm}{!}{\includegraphics{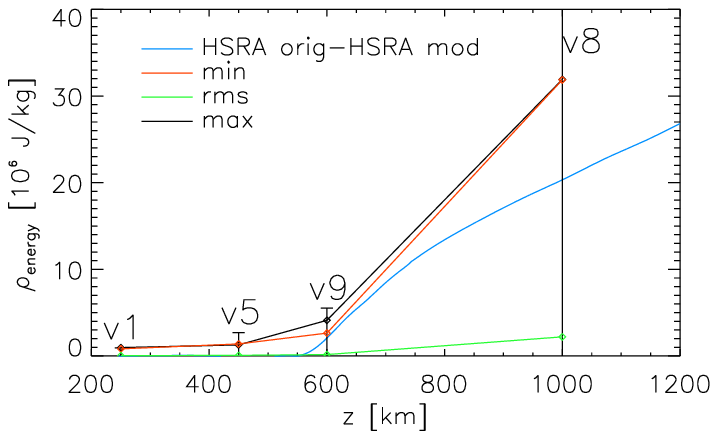}}}
\includegraphics{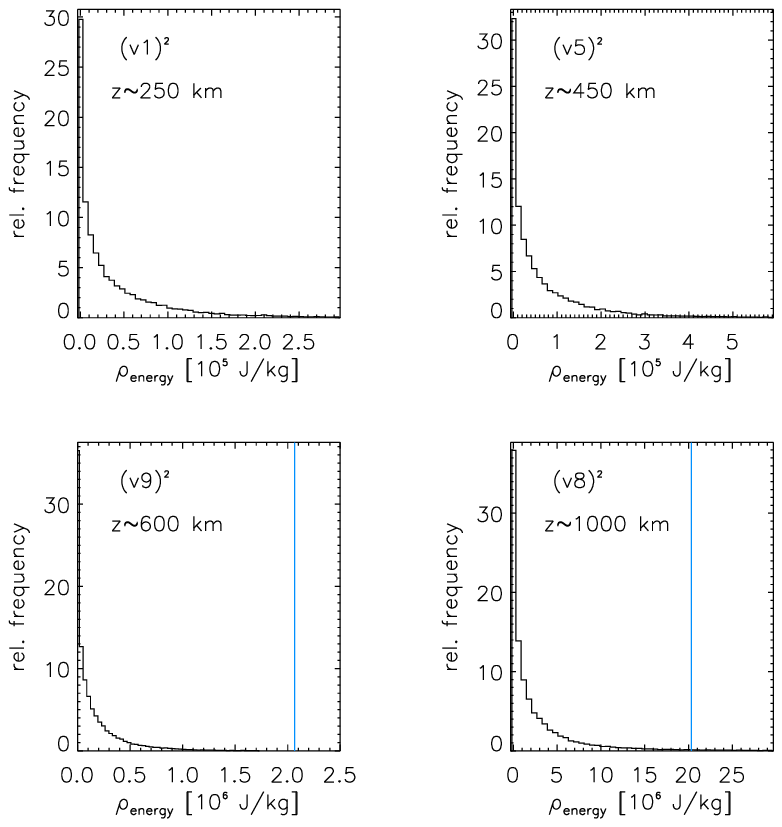}
\caption{({\em Top panel}): mass energy density, $E/\rho=1/2\,v_i^2$, vs height. {\em Red, green and black lines} correspond to minimal, rms, and maximal velocity, respectively. The {\em blue line} gives the enhancement of internal energy $\Delta p_{int}$ required to obtain the original HSRA temperature stratification.  The {\em vertical black lines} indicate the extension of the velocity distributions for each line that are shown in the {\em lower four panels}. There the {\em blue vertical lines} denote the energy required to obtain the HSRA model at the specific formation height of the line.\label{mass_energy_density}}
\end{figure}
\paragraph{Comparison to a semi-empirical atmosphere model} All velocities can be converted to a mechanical energy equivalent by
\begin{equation}
E_{mech} = 1/2\, \rho_{gas} \cdot v^2 \;. \label{eq_velo}
\end{equation}
More precisely, $E_{mech}$ is the ram pressure and can be considered as a mechanical energy {\em density}. To distinguish it from the mass energy density used below, we will refer to the quantity as ``volume energy density'' in the following.

\begin{figure}
\centerline{\resizebox{6.5cm}{!}{\includegraphics{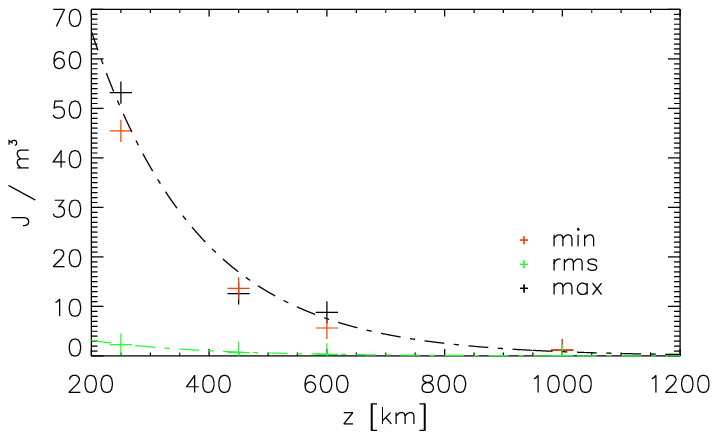}}}
\centerline{\resizebox{6.5cm}{!}{\includegraphics{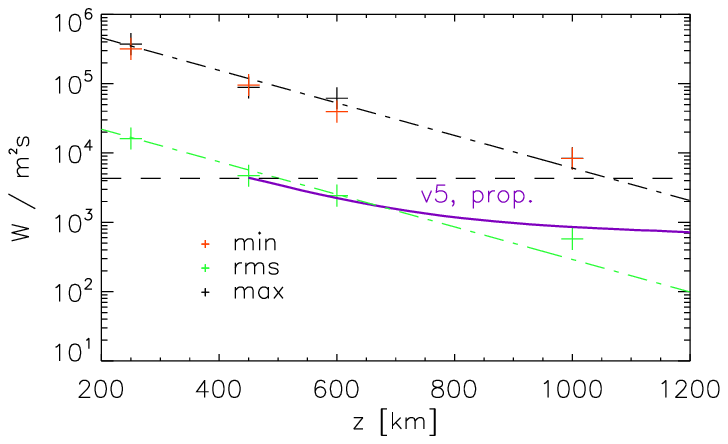}}}
\centerline{\resizebox{6.5cm}{!}{\includegraphics{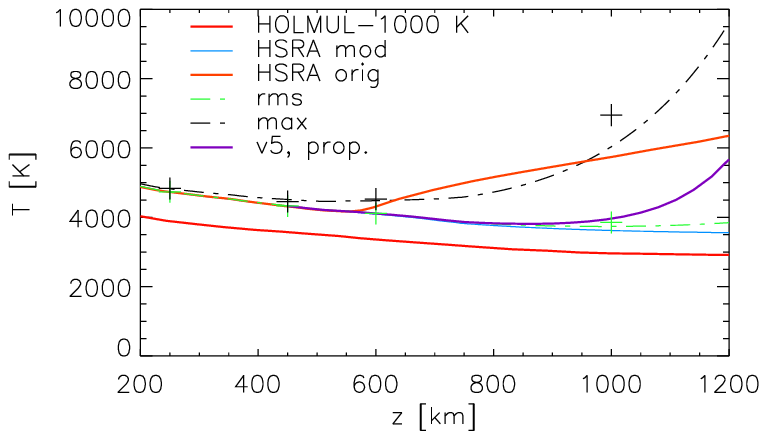}}}
\caption{Volume energy density, $E/V=1/2\,\rho\,v_i^2$, vs height ({\em top panel}). The {\em green} and {\em black dash-dotted} lines show a fit of an exponential decay. ({\em Middle panel}): acoustic energy flux in logarithmic scale, obtained as $E/V\cdot c_s$. The {\em purple} line shows the result of propagating v5 in height. The {\em dashed horizontal} line denotes a commonly used chromospheric energy requirement. The {\em lower panel} shows the temperature stratifications resulting from the dash-dotted and purple lines of the upper panel in comparison to the original (modified) HSRA and the HOLMUL atmosphere model. \label{volume_energy_density}}
\end{figure}

For the comparison with a semi-empirical solar atmosphere model, we chose the temperature enhancement over radiative equilibrium (RE) temperature as the quantity that should correspond to the volume energy density. RE conditions can be thought of as providing a lower limit to the temperature for the upper photosphere and low chromosphere on a long-term average because of the pervasive radiation field from continuum layers that constantly tries to re-heat plasma above to RE temperature \citep[e.g.][]{ulmschneider+etal1978,cheung+etal2007}. We selected the HSRA atmosphere model \citep{gingerich+etal1971}, because the differences between the various semi-empirical models \citep[e.g., in][]{gingerich+etal1971,vernazza+etal1981,fontenla+etal1999,fontenla+etal2006,avrett2007} on the location and strength of the chromospheric temperature rise are minor in the present context. To obtain a RE atmosphere model, we exchanged the temperature stratification of the HSRA atmosphere model with that of the Holweger-Mueller model \citep[HOLMUL,][]{holweger+mueller1974} for all optical depths smaller than log $\tau$=-4 (z$\sim$550 km). The HOLMUL model roughly corresponds to an atmosphere in RE condition (cf.~ Fig.~\ref{volume_energy_density} for a plot of all stratifications). As the semi-empirical atmosphere models like HSRA are assumed to correspond to a temporal and spatial average of atmospheric conditions, we will do the comparison to the velocity {\em distributions}, because their range should cover the average status predicted by the semi-empirical model. 

We convert the temperature enhancement over RE conditions, $\Delta T$, to an enhancement of the internal energy density, $\Delta p_{int}$ by
\begin{equation}
\Delta p_{int} = \rho_{gas} \cdot \frac{R \Delta T}{\mu (\gamma-1)} \;, \label{eq2}
\end{equation}
where $R = 8.31 J{\rm mol}^{-1}K^{-1}$, $\mu =  1.3 g{\rm mol}^{-1}$ and $\gamma = 5/3$ are the gas constant, the specific mass and the adiabatic coefficient, respectively.

For the comparison of $\Delta p_{int}$ with the mechanical energy of the waves, it actually is not necessary to take the gas density $\rho_{gas}$ into account. $\rho_{gas}$ in chromospheric layers usually comes with a large uncertainty: if the density is derived from the condition of hydrostatic equilibrium, it strongly depends on the assumed temperature stratification. One can, however, simply compare 1/2 $v^2$ with $R/ (\mu (\gamma-1)) \cdot \Delta T$. Both equations yield a (mass) energy density in Jkg$^{-1}$. Consider for example the Ca line-core velocity and the corresponding enhancement of HSRA over RE at 1000 km. $R/\mu(\gamma-1)$ is around 9600 and $\Delta T\sim$2000 K; thus, the internal energy enhancement is $9600 \cdot 2000 \sim  1.9\cdot 10^{7}$ Jkg$^{-1}$. This is to be compared with $0.5\,\sigma_{v8}^2 = 0.5\cdot 2100^2 \sim 2\cdot 10^{6}$ Jkg$^{-1}$ from the LOS velocity, which is an order of magnitude smaller.

Figure \ref{mass_energy_density} shows how the enhancement of the internal energy density required to maintain the original HSRA model above radiative equilibrium compares with the velocity statistics. For the lines 1 and 5, with a formation height below z$<$500 km, actually no additional energy is required to maintain the HSRA model. For lines 9 and 8, the extreme  velocities ({\em black line and red line}) would supply more than enough energy compared to the requirements of HSRA, but the rms velocities are clearly insufficient. This is highlighted in the distribution of the mass energy density in the {\em lower panels}. The energy required to obtain the HSRA model lies at the upper limit of the distributions for lines 9 and 8, and is far from the rms value. We point out that the mass energy density increases with height, reflecting the increase of the velocity amplitude.

When the gas density is multiplied to obtain a volume energy density, the trend with height is reversed. We used a gas density with height that corresponds to our RE temperature stratification put to hydrostatic equilibrium; the density is close to that of the HSRA model up to z=1000 km. As the histograms and the value required to obtain the HSRA would simply be scaled by the same coefficient, the same result as in Fig.~\ref{mass_energy_density} would result for the lower panels. We thus have used the plot of  volume energy density vs height ({\em  upper panel} of Fig.~\ref{volume_energy_density}) differently. We made a (manual) fit to the exponentially decreasing volume energy density with height to reproduce the observed values ({\em dash-dotted lines}). For both extreme and rms velocities, a reasonable agreement can be achieved with an exponential decay $\exp(-z/a)$ with a$\sim$185 km ({\em middle panel}). For easier comparison with other work, we converted the volume energy density here to an acoustic flux density by multiplying with a constant sound speed of $c_s\sim$ 7 kms$^{-1}$. We overplotted the curve that comes from propagating the velocity of line 5 upwards ({\em purple line}). The {\em horizontal dashed line} denotes 4.3 kWm$^2$s$^{-1}$ as a reference of chromospheric radiative losses \citep{vernazza+etal1976}; for z $>$ 500 km, the energy contained in the LOS velocities falls short of it. If the exponential decay curves of the upper panel are converted to the corresponding temperature enhancement over RE temperatures by equating Eqs.~(\ref{eq_velo}) and (\ref{eq2}), the extreme velocities would again be more than sufficient for the HSRA requirements, whereas the rms velocities would at least suffice for an increase of 500-1500 K above RE temperatures for heights above 1200 km.

Thus, in this comparison of mechanical energy contained in mass motions and the enhancement of internal energy required to lift an RE model to the original HSRA model, the mass motions are insufficient to supply  the necessary energy. Our interpretation of all line-core shifts as being due to mass motions corresponds to an upper limit, as some of the shifts of the blends inside the \ion{Ca}{II} H line may also actually have been due to temperature effects.
\section{Summary and discussion\label{summ_disc}}
We derived the mechanical energy contained in the LOS velocities of various spectral lines that form in different geometrical heights in the solar atmosphere. The lines could be grouped into four height ranges of which only the lowest two show a clear relation (anti-correlation) to the granulation pattern in the continuum intensity. When comparing the line-core velocities to the corresponding line-core intensities, a positive correlation is found for all lines but the lowest forming one: red-shifts are related to intensity increases. The amplitude of the velocity oscillations increases with height, both for the whole velocity distributions and co-spatial (but not co-temporal) velocities in a time-series, in good agreement with theoretical predictions from linear wave theory. When the wave energy is converted to a corresponding temperature enhancement of the internal energy, it falls short of the requirements of lifting a RE atmosphere model to the original HSRA model with a chromospheric temperature rise. The distributions of the LOS velocities barely reach these requirements for the most extreme, and thus rare, velocities. A fit of an exponential decay to the mechanical energy density with height yields a possible heating and a temperature increase in the chromospheric layers of around 500 to 1500 K, which is significantly less than in the HSRA atmosphere model.

{ \paragraph{Magnetic field influence on the \ion{Ca}{II} H line core}
The area fraction of \ion{Ca}{II} H profiles that show signatures of magnetic fields is around 15 \% (Sect.~\ref{mag_fields}). The rms velocities of all spectral lines considered are reduced on these locations, most prominent for the \ion{Ca}{II} H line core (Sect.~\ref{sec_velo}). The line-core velocity of Ca also shows a considerable contribution at low temporal frequencies ($\nu<$1.3 mHz) that is not present in this way for the other lines (Appendix \ref{appa}).
\begin{figure}
\resizebox{8.8cm}{!}{\hspace*{.5cm}\includegraphics{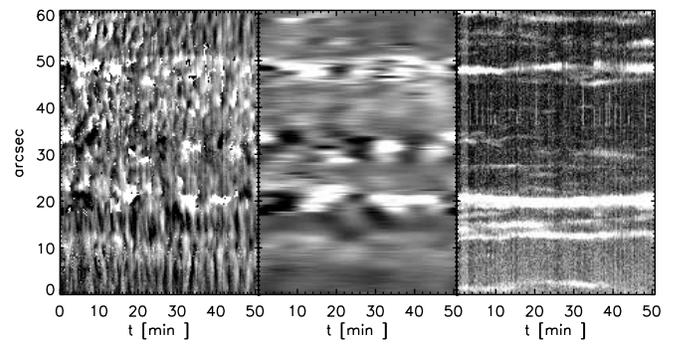}}\\$ $\\
\caption{{\em Left to right}: \ion{Ca}{II} H line-core velocity, low-frequency contributions, integrated absolute circular polarization signal. Velocities are displayed in a $\pm 5$ kms$^{-1}$ range.\label{low_freq}}
\end{figure}
This low-frequency contribution to the LOS velocity comes mainly from locations with photospheric magnetic fields \citep[Fig.~\ref{low_freq};][]{lites+etal1993}. The locations of the magnetic fields can nearly be determined from the Ca line-core velocity, by selecting the large-amplitude slow variations in Fig.~\ref{low_freq}. This is clear for the two persistent magnetic patches at y=20$^{\prime\prime}$ and 50$^{\prime\prime}$, but also at y=30$^{\prime\prime}$ a transient and weaker polarization signal is seen. Due to the small extent of the spatial scan in the time-series we cannot exclude that the polarization signal actually comes from a network element that the slit has intersected just at the very edge.

The importance of this field-related velocity component should increase with height in the atmosphere when the magnetic field lines spread out to fill the whole chromospheric volume. In the formation height of \ion{Ca}{II} H, only the locations with also photospheric polarization signal are affected which suggests that the Ca line in quiet Sun forms below the canopy layer that marks the transition from a basically field-free to fully magnetized atmosphere.}
\paragraph{Range of temperature variations}
If the velocity distributions are converted to their (relative) temperature equivalent, the extreme velocities of the Ca line core lead to a sufficient enhancement of around 2000 K above RE temperature to reach the HSRA temperature values. The distributions span the full range from 0 to the 2000 K, implying a variation of the temperature by several hundred K. For line 9 at around 600 km height, the temperature variation range already covers around 1000 K. This is at odds with the claim of \citet{avrett+etal2006} that temperature variations in the low chromosphere should not exceed $\pm 200$ K away from a ``standard'' temperature rise comparable to that in the HSRA. Their conclusion was based on the small amount of intensity variation in spectra observed with SUMER. \citet{avrett+loeser2008} repeat this claim by stating that ``{\em the brightness temperature variations observed at high spatial resolution, away from active regions, do not exceed a few hundred degrees}''. They also suggest strongly that any temperature fluctuations should be around a semi-static model with a temperature rise. We will address the question of the range of intensity fluctuations in the 2nd of the paper of this series in more detail, but remark already here that the intensity variations cover a similar dynamical range as the velocities and actually imply temperature fluctuations above 500 K around {\em a RE atmosphere without any chromospheric temperature rise}. \citet{rezaei+etal2008} have shown examples of \ion{Ca}{II} H profiles that cannot be reproduced with temperature stratifications including a ``standard'' chromospheric temperature rise, but actually most of the observed \ion{Ca}{II} H spectra disagree all the time with the assumption of a strong semi-static temperature rise that is only slightly modulated by dynamic events.
\paragraph{Relation between chromospheric intensity and LOS velocities} The line-core position of \ion{Ca}{II} H may be an unsuited tracer of the chromospheric dynamics due to the complex line formation. We thus use the intensity close to and in the line core of \ion{Ca}{II} H to discuss the causal relationship between chromospheric intensity and photospheric velocities. Most of the LOS velocity oscillations have temporal frequencies below the commonly assumed cut-off frequency of $\nu_c$ = 5 mHz (cf.~ Fig.~\ref{fig_fft} in Appendix \ref{appa}) and thus should not be able to reach chromospheric heights. \citet{beck+etal2008}, however, found that the phase shifts between the \ion{Ca}{II} H line-core intensity variations and intensity variations in the Ca line wing indicate propagating waves down to frequencies of around 2 mHz, the same as found by \citet{rutten+etal2004a} for G-band and Ca line-core intensities. The same applies to the phase shifts between Ca line-core intensity and photospheric LOS velocities (ibidem, their Fig.10 b). \citet{centeno+etal2006} and \citet{khomenko+etal2008} have shown that the presence of radiative energy losses in the atmosphere can lower the cutoff frequency. 
\begin{figure}
\resizebox{7.25cm}{!}{\hspace*{1.cm}\includegraphics{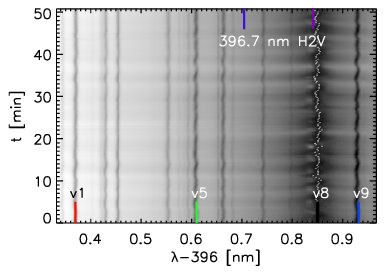}}\\$ $\\$ $\\
\centerline{\resizebox{7.cm}{!}{\includegraphics{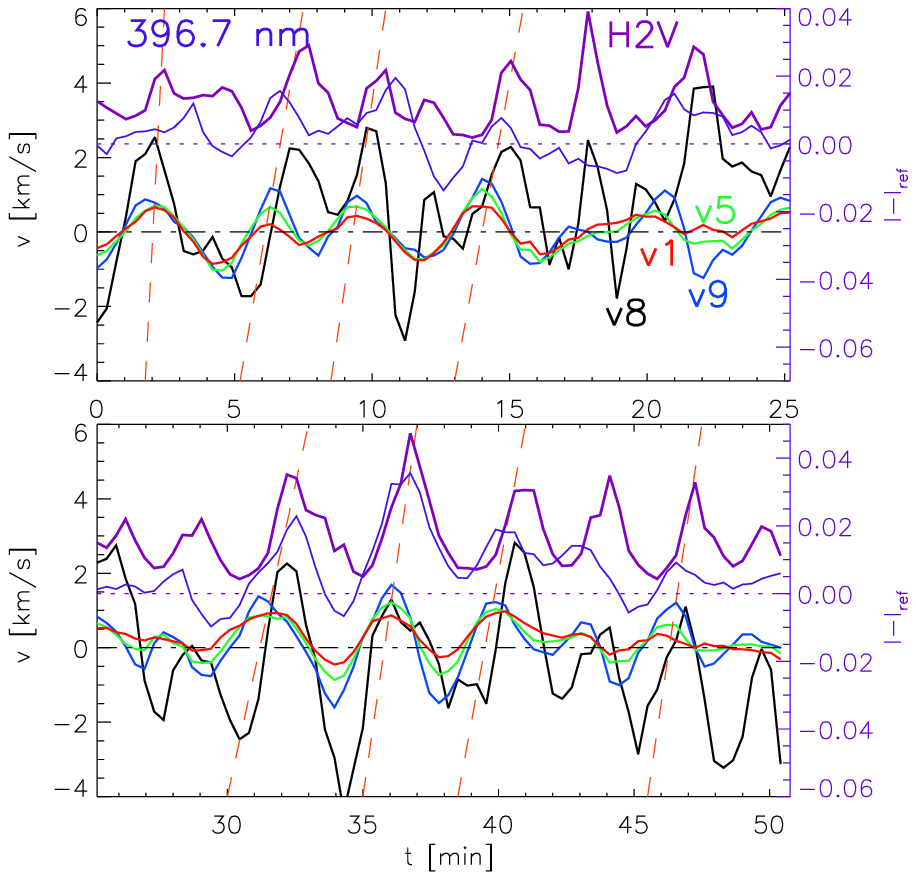}}}
\caption{{\em Top}: spectra at one fixed location along the slit with time. The {\em short vertical bars at bottom} denote lines 1,5,8 and 9, those {\em at top} 396.7 nm and the H$_{\rm 2V}$ emission peak. {\em Middle and bottom}: velocities and intensities from the spectra above for the time 0-25 min and 25-50 min, respectively. {\em Black, blue, red and green} lines show v8, v9, v5 and v1. The {\em thin blue line} shows the intensity at 396.7 nm, the {\em thick purple line} that of H$_{\rm 2V}$. {\em Dashed orange tilted lines} outline the propagation of perturbations with time.\label{no_timeshift}}
\end{figure}

That the frequency of the oscillations is of minor importance for the wave propagation, is also demonstrated in Fig.~\ref{no_timeshift}, where the same spectra and LOS velocities as used for Fig.~\ref{vel_exam} are displayed, but without applying any temporal shifts to obtain co-spatial velocities that are in phase. The position along the slit was at y=9\farcs3; Fig.~\ref{fig1} shows that there was a distance of at least 2-3$^{\prime\prime}$ to the nearest location with detected photospheric flux. %Note that \citet[][submitted]{vecchio+etal2008}, however, have shown recently that the photospheric field signal is insufficient for conclusions on the chromospheric field topology that inflcuences wave behavior up to a much larger distance from photospheric footpoints. 
In addition to the line-core velocities of lines 1 ({\em orange line}), 5 ({\em green}), 9 ({\em blue}), and 8 ({\em black}), we plotted also the intensity at 396.7 nm, around 0.15 nm distance from the Ca line core, and at the H$_{2V}$ emission peak ({\em upper two lines}). The intensity values were referenced vs the \ion{Ca}{II} H profile that resulted from the RE temperature stratification in a LTE calculation with the SIR code \citep{cobo+toroiniesta1992}. { The line-core velocity of \ion{Ca}{II} H, v8, was treated with a running mean over three time steps to reduce the scatter and improve the visibility of the velocity extrema.} 

It can be clearly seen in Fig.~\ref{no_timeshift} that in 90 \% of the cases an intensity increase of H$_{2V}$ is preceded by the same at 396.7 nm. The increase at 396.7 nm can again be traced back in time to a velocity oscillation of initially line 1. In some cases, the velocity oscillations of lines 1, 5, and 9 appear almost in phase (e.g., at t=31 min), but still trigger an increase of intensity at 396.7 nm and H$_{2V}$ with a time-lag, indicating a wave propagation with a photospheric origin (cf.~also Fig.~\ref{phase_diff}). The inclined {\em dashed lines} indicate the propagation  from a velocity oscillation of line 1 to an intensity increase of H$_{2V}$ in around 1 min. Even if most of the power is found below 5 mHz, still exactly these waves have a signature in the chromosphere in  H$_{2V}$ emission. If one would attribute a dominant period to the wave pattern of line 1 in Fig.~\ref{no_timeshift}, it would be closer to 5 min than 3 min { (see also Fig.~\ref{fig_fft})}. This is in some conflict with \citet{carlsson+stein1997} who found in their simulations of wave propagation driven by a photospheric piston that ``{\em runs with only low frequencies ($\nu<$4.7 mHz) do not produce grains}''. This could be reconciled with the present observations by the fact that the definition of a ``grain'' comes from the common usage of broad-band filter imaging and refers to stronger cases of shocks. The long-period oscillations of the velocity of line 1 lead to a transient intensity reversal in the Ca line core that would most probably not qualify as bright ``grain'' in most cases.

\paragraph{Energy flux at different heights} The energy flux with height of Fig.~\ref{volume_energy_density} shows a clear decrease throughout the whole height range. The absolute amount is very similar to the numbers given by \citet[][their Fig.~3]{straus+etal2008} who, however, applied a severe filtering in the Fourier space and used only the velocities corresponding to gravity waves. Similarly to the latter authors,  we also find that the mechanical energy at around z = 500 km is of the same order as the chromospheric radiative losses. If the formation height of line 9 is correctly assumed to be above 500 km, the mechanical energy seen in the velocity of this line, however, already falls below the common chromospheric energy requirements.

What amount of mechanical energy may be hidden from the observations ? In the time-series, one could invoke the temporal sampling that excludes all oscillations with periods below $\sim 40$ secs. The large-area map for comparison is a stochastic sample of waves at an arbitrary phase. The velocity histograms of both types of observations were found to be equivalent, thus, the temporal sampling of 21 secs does not seem to have an influence on the observed velocities. This leaves on the one hand the spatial resolution, and on the other the extension of the height layers that contribute to the respective line-core velocity. From the overview maps of the large-area scan it can be seen that the velocity patterns in the higher layers are spatially resolved, because they are more extended than the structures in the intensity maps; the typical size of structures also exceeds that of the clearly resolved granulation pattern. The typical width of intensity contribution functions in the Ca line wing is around 100-200 km (cf.~Fig.~\ref{response_geom}). Assuming that a similar height range contributes to each line-core velocity and a sound speed of $c_s=7$ kms$^{-1}$ leads to the fact that waves with wavelengths below $<$200 km, periods below 28 sec, and frequencies larger than 35 mHz cannot be observed. The power at frequencies above 20 mHz can be assumed to be negligible for chromospheric heating \citep[Fig.~\ref{fig_fft};][]{fossum+carlsson2005,carlsson+etal2007}, even if this was questioned by \citet{wedemeyer+etal2007} or \citet{kalkofen2007}. Note that all these authors did not consider oscillations below the cutoff frequency, whereas there are some indications that even these waves propagate vertically \citep[][or the present Fig.~\ref{no_timeshift}]{centeno+etal2006,beck+etal2008,khomenko+etal2008,straus+etal2008} and thus could contribute significantly to the chromospheric energy balance.

{ Another source of energy that could be missing in the observations are wave-related processes that have no line-of-sight component at disc centre. This will apply mainly to wave modes of the magnetic field lines. The field lines of strong photospheric flux concentrations are close to vertical, and several wave modes like sausage or kink modes have their main component perpendicular to the field lines. These waves then will only contribute to the statistical analysis of the LOS velocities with their small vertical component and remain otherwise undetected.}
\section{Conclusions}
\begin{figure}
\resizebox{8.8cm}{!}{\includegraphics{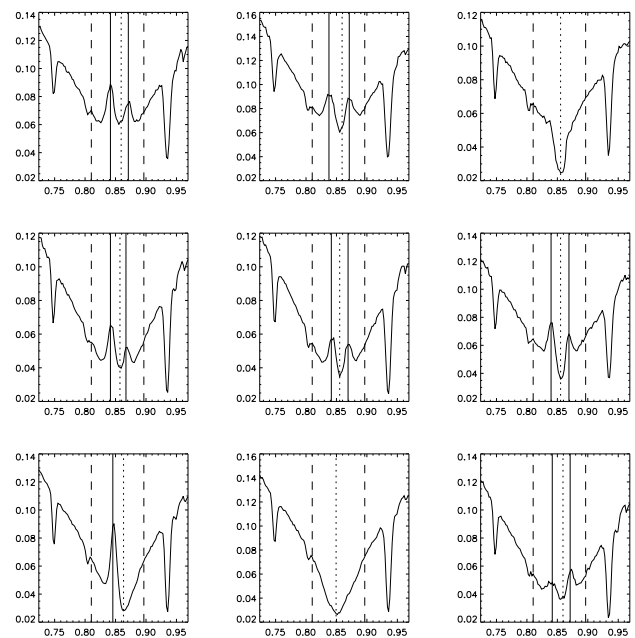}}
\caption{{ Examples of the analysis of the number of reversals in Ca II H core.  {\em Dashed lines} mark the analyzed wavelength range. {\em Dotted vertical line}: \ion{Ca}{II} H core position. {\em Solid vertical}: locations of intensity reversals. Wavelengths are $\lambda$-396 nm.}\label{fig_lobes}}
\end{figure}

We have analyzed the mechanical energy contained in line-of-sight velocity oscillations of spectral lines forming at different heights. The mechanical energy transported by the waves falls short of the requirements to lift a radiative equilibrium model atmosphere to one of the commonly used semi-empirical temperature stratifications with a strong chromospheric temperature rise. We find, however, that the chromospheric intensity variations are coupled to the LOS velocity of the lowermost forming spectral line, with a time-lag of around 1 min, and thus are triggered by photospheric motions. In the second paper of this series we will thus investigate, if the energy in the intensity variations, and in the intensity profiles as a whole, agree at all with the semi-empirical model.
\begin{acknowledgements}
R.R.~acknowledges support by the Deutsche Forschungsgemeinschaft under grant SCHM 1168/8-1. This research has been funded by the Spanish Ministerio de Educaci{\'o}n y Ciencia through project AYA2007-63881 and AYA2007-66502. The VTT is operated by the Kiepenheuer-Institut f\"ur Sonnenphysik (KIS) at the Spanish Observatorio del Teide of the Instituto de Astrof\'{\i}sica de Canarias (IAC). The POLIS instrument has been a joint development of the High Altitude Observatory (Boulder, USA) and the KIS.
\end{acknowledgements}
\bibliographystyle{aa}
\bibliography{references_luis_mod}
\begin{appendix}
{ \section{Analysis of \ion{Ca}{II} H profile shape\label{app_profiles}}
The \ion{Ca}{II} H line profile can show any number between zero or up to three intensity reversals near the line core. The number of reversals can be used for a qualitative estimate of which processes have led to the emission pattern. A symmetric double-peaked shape with equal intensity in  blue ($H_{\rm 2V}$) and red  ($H_{\rm 2R}$) emission peak indicates a static temperature rise in the absence of strong mass flows. Mass flows with vertical gradients lead to an asymmetry between $H_{\rm 2V}$ and $H_{\rm 2R}$ by shifting the absorption core. The absence of intensity reversals indicates a temperature stratification without a steep temperature rise in the {\em upper} atmosphere layers. This can either be a case of a monotonical decrease of temperature with height, but also happens during the passage of a shock wave when it still is located low in the atmosphere. A shock front lifts the intensity symmetrically to the line core until the reversals of $H_{\rm 2V}$ and $H_{\rm 2R}$ disappear, just before the onset of the strong emission phase \citep{liu1974,liu+skumanich1974, beck+etal2008}. 

We used a similar routine as \citet{rezaei+etal2008} to determine the number of emission peaks in the line core. The routine passes in wavelength through every profile, and searches for local maxima whose intensity exceeds that of the neighboring wavelength points. Due to the presence of noise, the spectra are treated with a running mean over five wavelengths points prior to the search for local maxima. Figure \ref{fig_lobes} shows nine randomly picked profile examples. The {\em solid vertical lines} mark the locations of the intensity reversals found in each profile. The {\em uppermost right} profile is a typical example for the presence of a shock front in the lower atmosphere (lifted wings at H$_{\rm 2V}$ and H$_{\rm 2R}$); the {\em lowermost left} one for a shock in the upper atmosphere (H$_{\rm 2V}$ emission). The {\em lowermost middle} panel shows a case of a reversal-free profile with a low H-index \citep[see][]{rezaei+etal2008}. }
\section{Fourier analysis and vertical propagation of velocity distributions\label{appa}}
\begin{figure}
\hspace*{.4cm}\resizebox{2.1cm}{!}{\includegraphics{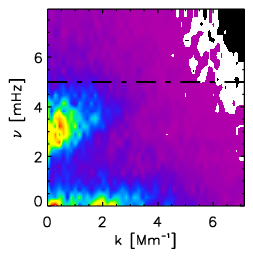}}\resizebox{2.1cm}{!}{\includegraphics{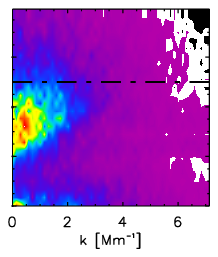}}\resizebox{2.1cm}{!}{\includegraphics{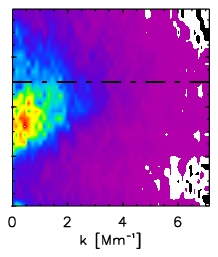}}\resizebox{2.1cm}{!}{\includegraphics{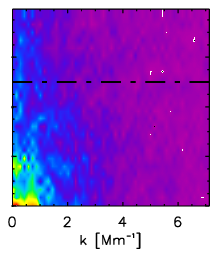}}\\
\caption{Power spectra of the LOS velocity of ({\em left to right}) line 1, 5, 9 and 8. The power is scaled between minimum and maximum in arbitrary units. The {\em dash-dotted line} denotes 5 mHz. \label{fig_fft}}
\end{figure}
\begin{figure}[hb!]
\resizebox{8.8cm}{!}{\hspace*{1cm}\resizebox{7cm}{!}{\includegraphics{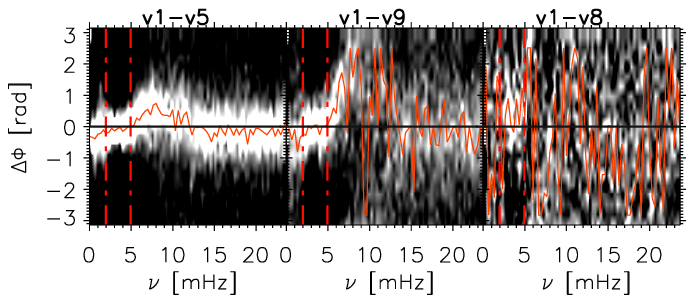}}\hspace*{.3cm}\resizebox{1.75cm}{!}{\includegraphics{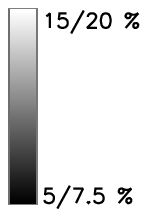}}}$ $\\$ $\\
\caption{Phase differences of LOS velocities as function of oscillation frequency. {\em Left to right}: v1 to v5, v1 to v9 and v1 to v8. The relative occurence of a phase difference value is given by the color code in the range of 5 \% to 20 \% (7.5 \%  to 15 \% for v1-v8). The {\em red curve} gives the center of gravity of the phase distribution at each frequency; {\em red vertical dashed lines} denote frequencies of 2 and 5 mHz, respectively.\label{phase_diff}}
\end{figure}
\begin{figure}
\resizebox{8.8cm}{!}{\hspace*{.5cm}\includegraphics{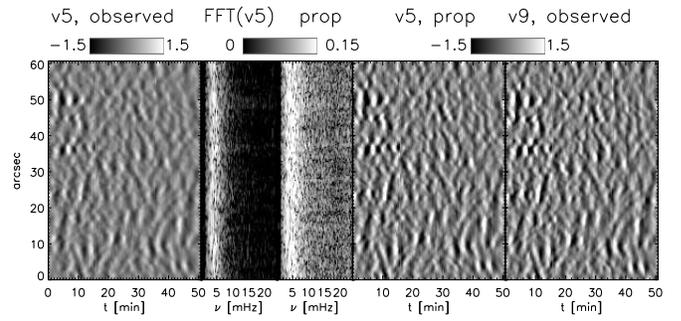}}\\$ $\\
\caption{Propagation of the velocity of line 5 to the height of line 9. {\em Left to right}: LOS velocity map of line 5, its Fourier transform, Fourier transform at the new height, velocity map at the new height, observed LOS velocity map of line 9. Velocities/amplitudes are in kms$^{-1}$ with the color coding at their top.\label{amp_prop}}
\end{figure}

\paragraph{Fourier power}
The time-series of \ion{Ca}{II} H spectra allows for a Fourier analysis to determine the dominant frequency contributions to the variations of the LOS velocities. Figure \ref{fig_fft} shows a k-$\nu$ diagram of the  unfiltered velocity maps of line 1,5, 9 and 8. The contribution at low frequencies ($<$ 1 mHz) is seen to diminish up to line 9, but is the strongest contribution for the line core of \ion{Ca}{II} H \citep[cf.~Fig.~\ref{low_freq} and][their Fig.~6]{rutten+etal2004a}. The power at medium frequencies ($\nu >$ 2 mHz) is seen to shift towards higher frequencies in lines 1,5, and 9. The greatest part of the power is located below the commonly assumed cut-off frequency of around 5 mHz ({\em dash-dotted horizontal line}). 
\paragraph{Phase differences}
In Fig.~\ref{phase_diff}, we show the phase difference between the lowermost forming line 1 and all others. The velocity variations show usually only a small or zero phase difference between the different lines, which would be expected if their frequencies are below the cut-off frequency. No clear relation with the line-core velocity of \ion{Ca}{II} H can be seen at all. The phase-relation between the \ion{Ca}{II} H line-core velocity and the low photospheric velocities shows a much larger scatter than the phase relation between the intensity of the H$_{2V}$ emission peak and photospheric velocities \citep[cf.][ Fig.~10b]{beck+etal2008}. This may be caused by the complex formation of the Ca line that renders the interpretation of the location of minimum line core intensity as a velocity doubtful { (see for example the lower left panel of Fig.~\ref{fig_lobes})}, and the fact that the main signature of the waves in the chromospheric layers is not a Doppler shift, but an intensity enhancement.
\paragraph{Propagation of velocity with height}
The velocities of the different lines can be propagated with height using the linear wave theory. The corresponding equation for the propagation contains a frequency-dependent amplitude scaling and is given by \citep{mihalas+mihalas1984}
\begin{eqnarray}
v(z+\Delta z,\omega) = v(z,\omega)\cdot \exp\left( \left(\frac{1}{2H} - \frac{\sqrt{\omega_c^2-\omega^2}}{c_s}\right)\cdot \Delta z \right)\label{scale_eq}\\ \mbox{for } \omega < \omega_c, \mbox{and by} \nonumber\\
v(z+\Delta z,\omega) = v(z,\omega)\cdot \exp\left(\frac{1}{2H} \cdot z\right) \;\mbox{for } \omega > \omega_c \;.\label{scale_eq1}
\end{eqnarray}
We used a constant scale height of $H$ = 100 km, a constant sound speed $c_s$ of 7 kms$^{-1}$ and an acoustic cutoff frequency of $\nu_c = 5$ mHz to calculate $\omega_c=2\pi \nu_c$. The frequency-dependent increase of the amplitude shifts the distributions of Fourier power towards higher frequencies. To obtain the velocity map corresponding to, e.g., the velocity of line 5 at the height of formation of line 9, one needs the contribution of each frequency to the velocity patterns in line 5, $ v(z_{v5},\omega)$. To this extent, we took the Fourier transform of each position along the slit in the time-series (Fig.~\ref{amp_prop}, {\em 2nd column}) of the velocity map of v5 ({\em 1st column}). The velocity amplitudes of the Fourier decomposition are then scaled up according to their frequency with Eqs.~(\ref{scale_eq}) and (\ref{scale_eq1}) ({\em 3rd column}). The result is transformed back to yield the propagated velocity map at the new height ({\em 4th column}). For comparison, the actual observed velocity map of line 9 is show in the {\em 5th column}. In a direct visual comparison the propagated and observed velocity map are identical. 

\begin{figure}
\resizebox{8.8cm}{!}{\includegraphics{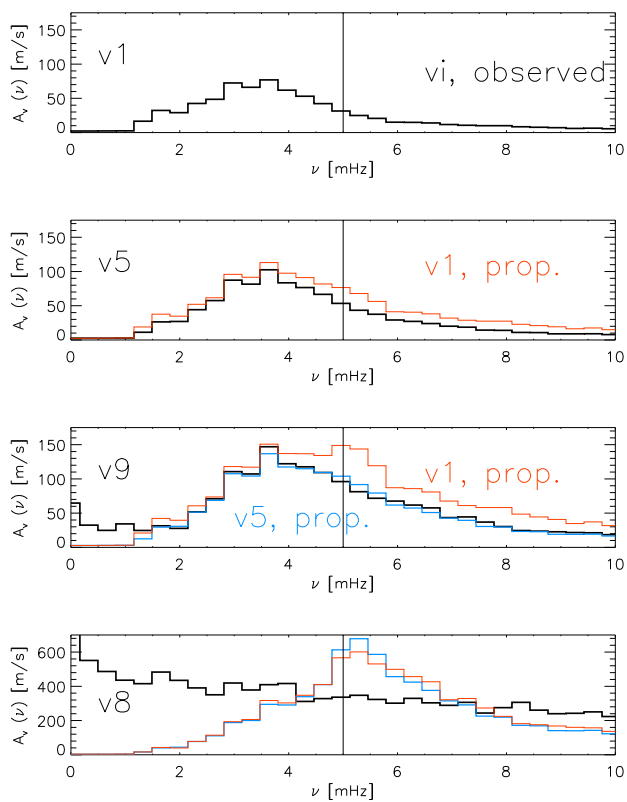}}
\caption{Velocity amplitudes of the Fourier transform, averaged along the slit. {\em Top to bottom}: line 1, 5, 9, and 8. {\em Black} lines shows the distribution in the observations, {\em blue} and {\em red} lines the one resulting from propagating line 1 and line 5, respectively.\label{fourier_dist}}
\end{figure}

This close agreement also holds for the distribution of the velocity power with frequency (cf.~Fig.~\ref{fourier_dist}). The propagated velocity of line 5 is again nearly indistinguishable from the observed one of line 9. For line 9 and for line 8 (\ion{Ca}{II} H line core), we used here the velocity maps without filtering for granular low frequency power. Line 8 shows a strong contribution from low frequencies that is not reproduced by the propagated velocities, even if the rms velocity values of both observed and propagated velocities turn out to be similar.
\end{appendix}

\end{document}